\documentclass[prd,aps,twocolumn,nofootinbib,superscriptaddress,eqsecnum,floatfix,preprintnumbers,amsmath,amssymb,
nofootinbib,longbibliography]{revtex4-2}

\usepackage{amssymb,amsmath}
\usepackage{epsfig}
\usepackage[dvipsnames]{xcolor}
\usepackage[utf8]{inputenc}
\usepackage{stmaryrd}
\usepackage{mathrsfs}
\usepackage{mathalfa}
\usepackage{accents}
\usepackage[normalem]{ulem}

\usepackage{graphicx}
\graphicspath{ {c:/Documents/PhysicsNotes/GradResearch/Images/GR_Lens/} }

\newcommand{\pa}{\partial}

\newcommand{\be}{\begin{equation}}
\newcommand{\ee}{\end{equation}}
\newcommand{\bea}{\begin{eqnarray}}
\newcommand{\eea}{\end{eqnarray}}
\newcommand{\ba}{\begin{equation}\begin{aligned}}
\newcommand{\ea}{\end{aligned}\end{equation}}

\newcommand{\beg}{\begin{gather*}}
\newcommand{\eng}{\end{gather*}}
\newcommand{\hh}{,\hspace{0.5cm}}
\newcommand{\hhh}{,\hspace{0.2cm}}

\newcommand{\n}[1]{\label{#1}}

\newcommand{\CAL}{\mathcal}

\newcommand{\ts}[1]{{\boldsymbol{#1}}}

\def\XXint#1#2#3{{\setbox0=\hbox{$#1{#2#3}{\int}$ }
\vcenter{\hbox{$#2#3$ }}\kern-.6\wd0}}

\usepackage[makeroom]{cancel}
\usepackage[caption=false]{subfig}
\usepackage[colorlinks=true,
            citecolor=green,
            linkcolor=red,
            filecolor=cyan,
            urlcolor=magenta,
            backref=false]{hyperref}

\begin{document}

\title{Spinoptics in the Schwarzschild spacetime}

\author{Valeri P. Frolov}%
\email[]{vfrolov@ualberta.ca}
\affiliation{Theoretical Physics Institute, Department of Physics,
University of Alberta,\\
Edmonton, Alberta, T6G 2E1, Canada}
\affiliation{
Center for Gravitational Physics and Quantum Information, Yukawa Institute for Theoretical Physics,
Kyoto University, 606-8502, Kyoto, Japan
}


\begin{abstract}
We study spinoptics equations in the Schwarzschild spacetime. We demonstrate that using the
explicit and hidden symmetries of this metric one can explicitly solve the equations for complex null tetrad associated with null rays representing photon's and graviton's motion. This allows one to integrate the spinoptics equations both for the electromagnetic and gravitational waves. It is shown that the main effect of the interaction of the spin of these fields with the spacetime curvature  is the tilt of the asymptotic planes of the massless particle orbit. The corresponding tilting angle is calculated. It is shown that this angle grows when a null ray passes in the vicinity of the circular null orbit located at $r=3M$.

...

 \hfill {\scriptsize Alberta Thy 8-24}
\end{abstract}

\maketitle

\section{Introduction} \label{s1}

Most information about the Universe is obtained by observing electromagnetic and (since September 2015) gravitational waves. Maxwell and gravitational wave equations near black holes and  in the cosmological background allow complete separation of variables and their solutions can be presented in the form of an infinite series. However, in practice using this representation for obtaining concrete results quite often is rather complicated. A powerful method of study high-frequency electromagnetic and gravitational waves is a well-known geometric optics approximation. The basic idea of this approach is that when the wavelength of the radiation is much less than other characteristic length parameters (such as radius of the beam of light, duration of the light pulse,  the curvature of the wavefront and the spacetime curvature), one can approximate a solution by assuming that locally it is similar to a monochromatic plane wave. Formally this is achieved by expanding the wave solutions in the inverse powers of the frequency and keeping a few lowest order terms of this expansion.

In a general case, the short wave (or high frequency) approximation is a powerful method of construction of approximate solutions of linear differential equations with spatially and time varying coefficients. It allows one to find asymptotic solutions by reducing this problem to study Hamiltonian dynamical systems. This method is widely used in different areas of physics and has different (historically motivated) names. In quantum mechanics this method is known as a quasiclassical or WKB approximation. In wave optics it is known as a geometric optics. It takes its origin with the paper by Debay in 1911
\cite{Debay:1911}.

In the application to the wave equation, one searches for a solution of the form $\Phi=a exp(i\omega S)$, where $\omega S$ is a fast changing function,  called an eikonal, and  $a$ is a slowly changing amplitude. The lowest order in 1/ expansion gives the Hamilton-Jacobi equation $H(x,\nabla S)$ for the Hamiltonian $H$ . The corresponding Hamilton equations determine a set of null rays with tangent vectors $l_{\mu}=S_{,\mu}$. Such rays form a Lagrangian submanifold in the phase space,  which can be used to find  the eikonal function for its given initial value.  The subleading term in the $1/\omega$ expansion gives a transport equation which allows one to find the amplitude $a$.

The application of the geometric optics to the propagation of the high frequency electromagnetic and gravitational waves in a curves spacetime gives the following results: (i) Light rays are null geodesics, (ii) Polarization vector is perpendicular to the rays and is parallel-propagated; (iii) The amplitude is governed by an adiabatic invariant which, in quantum language, states that the number of photons (or gravitons)  is conserved (see e.g. \cite{MTW}).

A gravitational field can cause a rotation of the polarization plane of light. This phenomenon is known as the gravitational Faraday effect  (see e.g. \cite{Skrotskii,Plebanski,God,GF2,GF3,GF4,GF5,Piran:1985,GF6,CariniRuffini,Perlick_1993,GF7,GF8,GF9,Halilsoy:2006ev,GF10,Frolov:2011mh,Ghosh,Dolan:2018nzc,Dolan:2018ydp,Hou:2019wdg,Li:2022izh,
Shoom:2022oer} and references therein). It arises due to interaction of spin of the light with the curvature. As a result of the spin-orbit interactions, the left- and right-handed circularly polarized waves  propagates slightly differently. Dependence of photon's trajectory in a gravitational field on its polarization is an analogue of optical Magnus effect (Hall effect of light). The spinoptics is a modification of the geometric optics approximation which allows one to take into account this effect\footnote{Let us note that the equations of motion of a massive rotating object  in an external gravitational field (Mathisson–Papapetrou–Dixon equations)  were obtained long time ago  \cite{MATHISSON,PAPAPETROU,DIXON} and have been studied in detail. However, these equations do not allow a well defined limit when the mass of the object  tends to zero.}. The spinoptics for the electromagnetic waves was discussed in \cite{Frolov:2011mh,Frolov:2012zn,Yoo:2012vv,Oancea:2019pgm,Oancea:2020khc,
Frolov:2020uhn,Dahal:2022gop}. Spinoptic approach for the gravitational waves in a curved spacetime was considered in \cite{Yamamoto:2017gla,Andersson:2020gsj,Dahal:2021qel,Kubota:2023dlz}.

A key observation in the spinoptics approach is a following. Let us consider a scattering of light by a massive object and assume that the emitter and detector are located at far distance $L$ from this body, while the light passes close near it. In this problem besides the small parameter of the wavelength $\lambdabar=1/\omega$ there exists a large length parameter $L$. In this situation even tiny spin dependent corrections to the equations of motion may become essential at large distance.
To obtain the equations of the spinoptics one  considers right and left circular polarized waves independently and develop high- frequency approximation for each of these states. Using these high-frequency expansions  one  searches for the helicity dependent $1/\omega$  terms and  includes these helicity sensitive corrections into the eikonal equations. Such a modification enhances  of helicity-dependent corrections and slightly changes the worldlines of the massless particles with spin. These worldlines still are null but  not geodesic anymore.

There exists two ways to derive the spinoptics equations:
\begin{itemize}
\item Substitute the ansatz for high-frequency circular polarized waves into
the field equations. After this collect  helicity
sensitive terms and  include them into the eikonal equation \cite{Oancea:2019pgm,Oancea:2020khc,Frolov:2020uhn}.
\item Substitute the ansatz for high-frequency circular polarized waves  into the field
action, and use the obtained effective action for deriving
spinoptics equations \cite{Frolov:2024ebe,Frolov:2024qow}.
\end{itemize}
Let us emphasize that the spinoptics equations obtained by both methods are the same. However, the second method is  simpler and more transparent  than the first one.

In a general case the spinoptic equations are rather complicated. Besides the equations for the propagation of the null rays and  scalar amplitude, there are also  additional equations for the propagation of the vectors of the complex null tetrads associated with the null rays. This set of equations can be studied numerically. However, there exists astrophysically interesting case when an analytic study of the spinoptics equations is  possible. This is a case of a black hole in an asymptotically flat spacetime. In this paper we discuss solutions of the spinoptic equations in the Schwarzschild spacetime and demonstrate how they can be solved. The Schwarzschild metric has four Killing vectors. As a result, the  geodesic equations for
particles and light are completely integrable and their orbits are planar. Moreover, the solutions of these equations can be written in an explicit form in terms of the elliptic integrals (for more discussion and references see e.g. \cite{FRO_ZEL}).
Besides Killing vectors responsible for the "explicit" symmetry  of the Schwarzschild spacetime it also possesses  a special, rank two, closed conformal Killing-Yano tensor  generating the "hidden" symmetry. We shall demonstrate how this hidden symmetry helps one to integrate the spinoptics equations.
We expect that a similar approach can be used for solving the spinoptics equations in the Kerr metric as well.

In section~II we collect  useful relations  for the Schwarzschild geometry, including expressions for the generators of the hidden symmetries, which are used later in the paper. In section~III we discuss null rays and complex null tetrads associated with them. Using an
approach similar to the method of Marck \cite{MARCK_2}, we solve equations of the parallel transport for such a tetrad when a null ray is a geodesic. Spinoptics equations and their solutions in the Schwarzschild geometry are discussed in section~IV. Section~V contains a brief summary of the obtained results and  discussion.

In this paper we use sign conventions of the book \cite{MTW} and geometric units of $c=G=1$. We also denote 4D objects, such as 4D vectors and tensors, by boldface symbols.

\section{Null rays in the Schwarzschild spacetime}

The Schwarzschild metric has the form
\be
\begin{split}
ds^2&=-f dt^2+\dfrac{dr^2}{f}+r^2 d\Omega^2\, ,\\
d\Omega^2&=d\theta^2+\sin^2\theta d\phi^2\, .
\end{split}
\ee
It admits  Killing vectors
\be
\xi^{\mu}=(1,0,0,0)\hh \zeta^{\mu}=(0,0,0,1)\, ,
\ee
and a closed conformal Killing-Yano tensor $h_{\mu\nu}$, which is an antisymmetric rank two tensor satisfying the equation
\be
h_{\mu\nu;\lambda}=\xi_{\mu}g_{\nu\lambda}-\xi_{\nu}g_{\mu\lambda}\, .
\ee
The form $\ts{h}$ is closed, $d\ts{h}=0$, and hence it has a potential  $\ts{h}=d\ts{b}$.
For the Schwarzschild geometry
\be
\begin{split}
&b_{\mu}=-\dfrac{1}{2}r^2\delta_{\mu}^{t}\, ,\\
& h_{\mu\nu}=2b_{[\mu ,\nu]}=-r(\delta_{\mu}^t \delta_{\nu}^r-\delta_{\mu}^r \delta_{\nu}^t)\, .
\end{split}
\ee
Its dual is the Killing-Yano tensor
\be
k_{\mu\nu}=\dfrac{1}{2}e_{\mu\nu\alpha\beta}h^{\alpha\beta}\, .
\ee
It obeys the equation
\be \n{kkk}
k_{\mu(\nu ;\lambda)}=0\, .
\ee
In the Schwarzschild metric this tensor has the following form
\be
k_{\mu\nu}=r^3 \sin\theta(\delta_{\mu}^{\theta}\delta_{\nu}^{\phi}-\delta_{\nu}^{\theta}\delta_{\mu}^{\phi})\, .
\ee

Using $k_{\mu\nu}$ one can define the symmetric rank two Killing tensor $K_{\mu\nu}$
\be
K^{\mu}_{\ \nu}=k_{\mu}^{\ \alpha}k_{ \nu\alpha}=r^2 [\delta^{\mu}_{\ \theta}\delta_{\nu}^{\ \theta}+\delta^{\mu}_{\ \phi}\delta_{\nu}^{\ \phi}]\, .
\, .
\ee
We also define other  rank two tensors
\be
H^{\mu}_{\ \nu}=h^{\mu\alpha}h_{\alpha \nu}\hhh
\tilde{K}_{\mu\nu}=h^{\mu\alpha}k_{\alpha \nu}\hhh
\tilde{H}_{\mu\nu}=k^{\mu\alpha}h_{\alpha \nu}\, .
\ee
Simple calculations give
\be\n{HHKK}
\begin{split}
&H^{\mu}_{\ \nu}= r^2 [\delta^{\mu}_{\  t}\delta_{\nu}^{\ t}+\delta^{\mu}_{\ r}\delta_{\nu}^{\ r}  ]\, ,\\
&\tilde{K}^{\mu}_{\nu}=\tilde{H}^{\mu}_{\nu}=0\, .
\end{split}
\ee

Consider a vector $u^{\mu}$ and use it define two new vectors
\be
u_k^{\mu}=k^{\mu}_{\ \nu} u^{\nu}\hh
u_h^{\mu}=h^{\mu}_{\ \nu} u^{\nu}\, .
\ee
Since tensors $\ts{k}$ and $\ts{h}$ are antisymmetric, both these vectors are orthogonal $u^{\mu}$
\be
u_{\mu}u_k^{\mu}=u_{\mu}u_h^{\mu}=0\, ,
\ee
while relations \eqref{HHKK} imply that they are also mutually orthogonal. In the next section we use this observation for construction of the complex null tetrad associated with null rays.

\section{Null rays and complex null tetrad associated with them}

Consider  a null geodesic $x^{\mu}=x^{\mu}(\lambda)$  in the Schwarzschild spacetime and let $\lambda$ be an affine parameter. The tangent vector  to this worldline can be written in the form
\be \n{lll}
\begin{split}
l^{\mu}&=\dot{x}^{\mu}=\big(\dfrac{E}{f},\dfrac{\CAL{R}}{r}, \dfrac{\Theta}{r^2 \sin\theta},\dfrac{L_z}{r^2\sin^2\theta}\big)\, ,\\
\CAL{R}&=\pm \sqrt{E^2r^2-L^2f}\hh
\Theta=\pm \sqrt{L^2\sin^2\theta-L_z^2}\, .
\end{split}
\ee
Here and later a dot means a derivative with respect to affine parameter $\lambda$.  For a null geodesics
the following quantities
\be
E=-\xi_{\mu}l^{\mu}\hhh L_z=\zeta_{\mu} l^{\mu}\hhh L^2=K_{\mu\nu} l^{\mu} l^{\nu}\, ,
\ee
are integrals of motion which have the meaning of the energy, azimuthal angular momentum and the square of the total angular momentum, respectively.
Relation \eqref{lll} shows that at a regular  point $(t,r,\theta,\phi)$ of the spacetime there exists only  one null geodesic with parameters   $E$, $L_z$, and $L^2$ passing through it. Signs $\pm$ which enter the $r$ and $\theta$ components of  $l^{\mu}$ are independent. They have the value $+1$ if in the motion of light the corresponding coordinates increases, and their value is $-1$ in the opposite case. The change of signs occurs at the corresponding turning points, where either $\CAL{R}$ or $\Theta$ vanish.
Using relation
\be
\dot{r}=\dfrac{dr}{d\lambda}=\dfrac{\CAL{R}}{r}
\ee
one obtains
\be
\lambda=\lambda_0+\int_{r_0}^r \dfrac{r dr}{\CAL{R}}\, .
\ee

Our goal now is to construct a complex null tetrad associated with a null geodesic ray.
For this purposes we define two vectors
\be\n{E12_0}
\tilde{e}_1^{\mu}=\dfrac{1}{L}h^{\mu}_{\ \nu}l^{\nu}\hhh
{e}_2^{\mu}=\dfrac{1}{L}k^{\mu}_{\ \nu}l^{\nu}\, .
\ee
As it was shown earlier,  these vectors are orthogonal to $l^{\mu}$ and are mutually orthogonal. The factor $1/L$ in the definition of these vectors is chosen so that they satisfy the following normalization conditions
\be \n{NORM}
\begin{split}
&\tilde{e}_1^{\mu}\tilde{e}_{1\mu}=e_2^{\mu}e_{2\mu}=1\, ,\\
&\tilde{e}_1^{\mu}e_{2\mu}=\tilde{e}_1^{\mu}l_{\mu}=e_2^{\mu}l_{\mu}=0\, .
\end{split}
\ee
In the Schwarzschild spacetime the vectors $\tilde{e}_1^{\mu}$ and $e_2^{\mu}$ have  the following components
\be\n{E12}
\begin{split}
&\tilde{e}_1^{\mu}=\big(\dfrac{\CAL{R}}{Lf},\dfrac{Er}{L},0,0 \big)\, ,\\
&e_2^{\mu}=\big( 0,0,\dfrac{L_z}{Lr\sin\theta},-\dfrac{\Theta}{Lr\sin^2\theta} \big) \,  .
\end{split}
\ee

To obtain the fourth vector of the tetrad we proceed as follows. Vectors $\tilde{e}_1^{\alpha}$ and  ${e}_2^{\beta}$ determine a 2D surface. We denote
\be
{\Sigma}_{\mu\nu}=e_{\mu\nu\alpha\beta}\tilde{e}_1^{\alpha} {e}_2^{\beta}\, .
\ee
The 2D plane determined by bi-vector ${\Sigma}_{\mu\nu}$ is orthogonal to the plane spanned by $\tilde{e}_1^{\alpha}$ and  ${e}_2^{\beta}$. It contains two null vectors which are the eigenvectors of $\Sigma_{\mu\nu}$. One of them, with the eigenvalue $-1$, is the vector $l^{\mu}$. We denote the other one, with the eigenvalue $+1$, by $\tilde{e}_3^{\mu}$. It has the form
\be\n{E3}
\tilde{e}_3^{\mu}=\big(  \dfrac{Er^2}{2fL^2},\dfrac{r\CAL{R}}{2L^2},-\dfrac{\Theta}{2L^2\sin\theta}, -\dfrac{L_z}{2L^2\sin^2\theta}
\big)\, .
\ee
By construction, this vector is orthogonal to vectors $\tilde{e}_1^{\alpha}$ and  ${e}_2^{\beta}$.
We choose $\tilde{e}_3^{\mu}$ to be future directed and normalize it by the condition
\be
\tilde{e}_3^{\mu} l_{\mu}=-1\, .
\ee

We use the constructed four vectors $l^{\mu},\tilde{e}_1^{\mu},{e}_2^{\mu},\tilde{e}_3^{\mu}$ to define a complex null tetrad parallel propagated along the null geodesic. Let us note that vector $l^{\mu}$ is parallel propagated by definition.
Let us show that the vector ${e}_2^{\mu}$ is also parallel propagated along the null geodesic. One has
\be
l^{\alpha}{e}_{2 \mu ;\alpha}=\dfrac{1}{L}\big(k_{\mu\nu;\alpha}l^{\alpha}l^{\nu}+k_{\mu\nu}l^{\alpha}l^{\nu }_{ \ ;\alpha}\big)=0\, .
\ee
The last equality for a geodesic null ray follows from the definition of the Killing-Yano tensor \eqref{kkk}.

Similarly, one has the following relation for the vector $\tilde{e}_1^{\mu}$
\be
l^{\alpha}\tilde{e}_{1 \mu ;\alpha}=\dfrac{1}{L}\big(h_{\mu\nu;\alpha}l^{\alpha}l^{\nu} +h_{\mu\nu}l^{\alpha}l^{\nu}_{\  ;\alpha}\big)=-\dfrac{1}{L}(l_{\alpha}\xi^{\alpha}) l_{\mu}\, .
\ee
Thus, the vector $\tilde{e}_1^{\mu}$ is not parallel propagated. However, it can be "improved" by adding to it a vector proportional to $l^{\mu}$
\be
{e}_1^{\mu}=\tilde{e}_1^{\mu}-\Phi l^{\mu}\, .
\ee
One can check that if the function $\Phi$ satisfies the equation
\be
\dfrac{d\Phi}{d\lambda}\equiv l^{\alpha}\Phi_{,\alpha}=-\dfrac{1}{L}l_{\alpha}\xi^{\alpha}=\dfrac{E}{L}\, ,
\ee
then the vector ${e}_1^{\mu}$ is parallel propagated along the ray\footnote{
A similar method for construction of the couple of vectors which are parallel propagated along null and timelike geodesics  in the Kerr geometry was described by Marck \cite{MARCK_1,MARCK_2}. See also \cite{LIVING}.
}.
A solution of this equation is as follows
\be
\Phi=\Phi_0+\dfrac{E}{L}\int_{r_0}^r dr \dfrac{r}{\CAL{R}}=\dfrac{E}{L}(\lambda-\lambda_0)+\Phi_0 \, ,
\ee

Using the constructed vectors we define the following complex null tetrad $(\ts{l},\ts{m},\bar{\ts{m}},\ts{n})$
where $\ts{l}$ is given by \eqref{lll} and the other vectors are defined as follows
\be\n{mn}
\begin{split}
&m^{\mu}=\dfrac{1}{\sqrt{2}}(e_1^{\mu}+i e_2^{\mu})\, ,\\
&n^{\mu}=\tilde{e}_3^{\mu}-\Phi \tilde{e}_1^{\mu}+\dfrac{1}{2}\Phi^2 l^{\mu}\, .
\end{split}
\ee
Here and later a bar denotes a complex conjugation.
One can check that the vector $\ts{n}$ is parallel propagated along the ray.
These vectors are normalized as follows
\be
l_{\mu}n^{\mu}=-1\hh m_{\mu}\bar{m}^{\mu}=1\, ,
\ee
the other scalar products vanish. We call set of vectors $(\ts{l},\ts{m},\bar{\ts{m}},\ts{n})$ a complex null tetrad associated with the null ray.

\section{Spinoptics in Schwarzschild metric}

\subsection{Spinoptics equations}

We discuss now the spinoptics equations in the Schwarzschild spacetime. These equations have the form \cite{Frolov:2020uhn,Frolov:2024ebe,Frolov:2024qow}
\be\n{DYNTET}
\begin{split}
&Dl^{\mu}=w^{\mu}\hh w^{\mu}=
\dfrac{\sigma}{\omega}(\bar{\kappa} m^{\mu}+\kappa \bar{m}^{\mu})\, ,\\
&Dn^{\mu}=0\hhh Dm^{\mu}=\dfrac{\sigma}{\omega} \kappa n^{\mu}\, ,\\
&\kappa =i R_{\mu \nu\alpha\beta} m^{\mu} l^{\nu} m^{\alpha}\bar{m}^{\beta}\, ,\\
&(a^2 l^{\mu})_{;\mu}=0\, .
\end{split}
\ee
Here $\sigma$ is the helicity which takes the value $\pm 1$ for photons and $\pm 2$ for gravitons.
In the leading order, that is in the limit $\omega\to \infty$, these equations describe null geodesic rays with a parallel propagated complex null tetrad. We consider the terms standing in the right-hand side of the equations \eqref{DYNTET} as perturbations and use spinoptics equations to find how this perturbation affects the null rays. Namely, we start with   a null geodesic with a tangent vector $\ts{l}_0$  and uses the relations \eqref{lll}, \eqref{E12}, \eqref{E3} and \eqref{mn} to calculate the complex null tetrad $(\ts{l}_0,\ts{m},\bar{\ts{m}},\ts{n})$ associated with $\ts{l}_0$. Using this tetrad one calculates the quantity $\kappa$ which enters the right-hand side of
equations  \eqref{DYNTET}. Since the parameter $\kappa/\omega$ is small, using the unperturbed tetrad  for the calculation of $\kappa$ is sufficient for calculation of the first order correction to the  null ray in the adopted approximation.

Using the expressions for $(\ts{l}_0,\ts{m},\bar{\ts{m}},\ts{n})$ derived in the previous section one  obtains a following expression for $\kappa$
\be
\kappa=\dfrac{3\sqrt{2}i}{2r^5}M L^2 \Phi\, .
\ee
One can also check, that the vector $w^{\mu}$ in the equations \eqref{DYNTET} is of the form
\be \n{wwee}
w^{\mu}=\dfrac{\sigma}{\omega}\dfrac{3ML^2\Phi}{r^5} e_2^{\mu}\, .
\ee

Let us note that the affine parameter is not uniquely defined and its choice allows a rescaling $\lambda\to A\lambda$.  It is convenient to perform the following scaling transformation $\hat{\lambda}=E\lambda$. Similarly, we rescale the other parameters. We use  a "hat" over the objects for their values after this rescaling. One has
\be
\begin{split}
&\hat{\lambda}=E\lambda\hhh\hat{l}^{\mu}=\dfrac{1}{E}l^{\mu}\hhh \hat{w}^{\mu}=\dfrac{1}{E^2}w^{\mu} \, ,\\
& \ell\equiv \hat{L}=\dfrac{L}{E}\hhh \ell_z\equiv \hat{L}_z=\dfrac{L_z}{E} \, ,\\
& \hat{\Phi}=E\Phi=\dfrac{1}{\ell}(\hat{\lambda}-\hat{\lambda}_0)+\Phi_0\, ,\\
&\hat{\CAL{R}}=\dfrac{1}{E}{\CAL{R}}=\pm \sqrt{r^2-\ell^2 f}\, ,\\
&\hat{\Theta}=\dfrac{1}{E}{\Theta}=\pm \sqrt{l^2\sin^2\theta-\ell_z^2}
\, .
\end{split}
\ee
To simplify the formulas, in what follows we  omit "hats". This means that in order to use the earlier obtained formulas it is sufficient to put there $E=1$ and substitute $L$ and $L_z$ by $\ell$ and $\ell_z$, respectively.
For the light ray scattering by the black hole the  quantity $\ell$ is  the impact parameter of the ray. In this "new" parametrization the equation \eqref{wwee} takes the form
\be\n{wwe2}
w^{\mu}=\psi e_2^{\mu}\hh
\psi=\sigma\lambdabar\dfrac{3M\ell^2\Phi}{r^5}
\, .
\ee
We use here the wavelength $\lambdabar=1/\omega$ instead of the frequency $\omega$.

For the problem under the consideration the calculations can be further simplified.
For a unperturbed geodesic motion in a spherically symmetric geometry the orbits are planar. Without loss of generality one can assume that a chosen null ray $\ts{l}_0$ lies in the equatorial plane. For this choice
\be
\theta=\pi/2\hhh \ell_z=\ell\hhh \Theta=0\hhh e_2^{\mu}\pa_{\mu}=\dfrac{1}{r}\dfrac{\pa}{\pa \theta}
\, .
\ee
The unperturbed null ray orbits are defined by the following equations
\be
\dfrac{dt}{d\lambda}=\dfrac{1}{f}\hhh
\dfrac{dr}{d\lambda}=\dfrac{\CAL{R}}{r}\hhh
\dfrac{d\phi}{d\lambda}=\dfrac{\ell}{r^2}\, .
\ee

For our choice, when $\theta=0$ corresponds to the "north pole" of the black hole, the vector $\ts{e}_2$ is directed
down with respect to the equatorial plane, in the direction of the increase of the angle $\theta$. Relation \eqref{wwe2} implies that the photons and gravitons with positive helicity during there motion near the black hole  will be slightly  deflected in the down direction with respect to the equatorial plane, while the particles with negative helicity are deflected in the opposite direction.

\subsection{Solving the equation}

 To study how the helicity affects the motion of a photon or graviton we write the vector of their 4D velocity in the form
\be\n{lPsi}
l^{\mu}=l_0^{\mu}+\Psi e_2^{\mu}\, .
\ee
Here $l_0^{\mu}$ the unperturbed 4D velocity for the geodesic motion and $\Psi e_2^{\mu}$ is a perturbation due to the spin-curvature interaction. To find $\Psi$ we shall use the spinoptics equatoions. In this approximation $\Psi=O(1/\omega)$. Since the vector $e_2^{\mu}$ is orthogonal to $l_0^{\mu}$, one has $l_{\mu}l^{\mu}=O(1/\omega^2)$. This means that in an adopted approximation the vector $l_{\mu}l^{\mu}$ is the null vector.

 Let us denote $D=l^{\mu}_0\nabla_{\mu}$, then one has
\be \n{pert}
Dl^{\mu}=D\Psi e_{2}^{\mu}\, .
\ee
Using equations \eqref{DYNTET} and \eqref{wwe2} one obtains
\be\n{PSI}
\dfrac{d\Psi}{d\lambda}=\psi\, .
\ee

We consider now an application of the obtained results to the scattering of light by a non-rotating black hole in the spinoptics approximation. We assume that a null geodesic ray is sent from the infinity with the some impact parameter $\ell$. In the absence of the spinoptics corrections it moves along a null geodesic in the equatorial plane.
We assume that the parameter $\ell$ is large enough so that the null ray passes  near the black hole at the some minimal radial distance $r_m>r_0$ and after this propagates back to the infinity. We choose coordinate $\phi$ so that  $\phi=0$ at  $r=r_m$. We also put the affine parameter $\tilde{\lambda}$ equal to zero at this point. Our goal is to find how the shift function $\Psi$ changes during this motion. Since $\psi$ depends on $r$, one should integrate \eqref{PSI} together with the radial equation for the unperturbed null geodesic.

Let us denote
\be
Z=\sqrt{1-\dfrac{\ell^2}{r^2}f}\, .
\ee
A turning point $r=r_m$  of the unperturbed null ray for a given impact parameter $\ell$ is defined by condition
\be\n{lrm}
\dfrac{\ell^2}{r_m^2}(1-\dfrac{2M}{r_m})=1\, .
\ee
To find a shift function $\Psi$ one can use the following set of the first order equations
\be\n{eq1}
\begin{split}
& \dfrac{dr}{d\lambda}=\pm Z\, ,\\
&\dfrac{d\Psi}{d\lambda}=\dfrac{A\ell \lambda}{r^5}\hh
A=3\sigma\lambdabar M
\, .
\end{split}
\ee
Here sign "minus"  in the first equation is used for the incoming part of the trajectory, while the sign "plus" is used for the outgoing part.
It is convenient to choose the  initial conditions at the turning point. However, the first of the equations \eqref{eq1} is singular at it. We use the following standard trick to overcome this problem. Instead of the first order equation for $r=r(\lambda)$ we use its second order version, which is obtained by its differentiation over $\lambda$.
The obtained set of equations  is
\be\n{eq2}
\begin{split}
& \dfrac{d^2r}{d\lambda^2}=\dfrac{\ell^2}{r^3}\big(1-\dfrac{3M}{r}\big)\, ,\\
&\dfrac{d\Psi}{d\lambda}=\dfrac{A\ell \lambda}{r^5}
\, .
\end{split}
\ee
We chose the initial conditions in the form
\be
r|_{\lambda=0}=r_m\hh  \dfrac{dr}{d\lambda}\big|_{\lambda=0}=0\hh \Psi|_{\lambda=0}=0\, .
\ee

One can further simplify the set of equations \eqref{eq2} by writing them in the dimensionless form. For this purpose we use the gravitational radius $2M$ as the standard length scale and change the parameters which enter the equations as follows
\be
\begin{split}
&\tilde{r}=r/(2M)\hhh \tilde{\ell}=\ell/(2M)\hhh \tilde{\lambda}=\lambda/(2M)\, ,\\
& \tilde{\lambdabar}=\lambdabar/(2M)\hhh \tilde{A}=A/(2M)^2=
\dfrac{3}{2}\sigma\tilde{\lambdabar}\hhh
\tilde{\Psi}={\Psi}/{\tilde{A}}
\,  .
\end{split}
\ee
The corresponding dimensionless form of the  equations is
\be\n{eq3}
\begin{split}
& \dfrac{d^2\tilde{r}}{d\tilde{\lambda}^2}= \dfrac{\tilde{\ell}^2}{\tilde{r}^3}\big(1-\dfrac{3}{2\tilde{r}}\big)\, ,\\
&\dfrac{d\tilde{\Psi}}{d\tilde{\lambda}}=\dfrac{\tilde{\ell} \tilde{\lambda}}{\tilde{r}^5}\, ,\\
&\dfrac{d\phi}{d\tilde{\lambda}}=\dfrac{\tilde{\ell}}{\tilde{r}^2}\, ,\\ &\dfrac{d\tilde{t}}{d\tilde{\lambda}}=\dfrac{\tilde{r}}{\tilde{r}-1}\, .
\end{split}
\ee
For completeness, we added here two unperturbed equations for the evolution of $\tilde{t}$ and $\phi$ coordinates.

The dimensionless impact parameter $\tilde{\ell}$ is related to $\tilde{r}_m$ as follows
\be
\tilde{\ell}=\dfrac{\tilde{r}_m^{3/2}}{\sqrt{\tilde{r}_m-1}}\, .
\ee
It is easy to check that $\tilde{\ell}$ has a minimum equal to $\tilde{\ell}_0=3\sqrt{3}/2$ at $\tilde{r}_0=3/2$.
This is an unstable circular orbit of the null ray $r=r_0$. For the impact parameter $\tilde{\ell}$ slightly larger that $\tilde{\ell}_0$ the unperturbed ray revolves around the black hole before it escapes to infinity.

Equation for $\phi=\phi(\tilde{\lambda})$ in \eqref{eq3} is invariant under simultaneous change of signs of $\tilde{\lambda}$ and $\phi$ and this coordinate monotonically increases from its asymptotic value $\phi=-\phi_{\infty}$ at $\tilde{\lambda}=-\infty$ to $\phi=\phi_{\infty}$ at $\tilde{\lambda}=\infty$. In the domain far from the black hole the incoming and outgoing rays  move practically along the straight line parallel to the radial line with the angles $\phi=\pm \phi_{\infty}$ and at the distance $\ell$ from them. We also  put $\tilde{\Psi}=0$ at the turning point.
Then this function monotonically increases from some asymptotic value $-\tilde{\Psi}_{\infty}$ to  $\tilde{\Psi}_{\infty}$.

Integration of a system of the first three equations in \eqref{eq3} allows one to obtain quantities $\phi$, $\tilde{\Psi}$ and $\tilde{r}$ as functions of $\tilde{\lambda}$. The plots of $\phi(\tilde{\lambda})$, $\tilde{\Psi}(\tilde{\lambda})$ for different choices of the parameter $\tilde{r}_m$ are shown at Fig.~\ref{F1} and Fig.~\ref{F2}, respectively.

\begin{figure}[!hbt]
    \centering
      \includegraphics[width=0.4\textwidth]{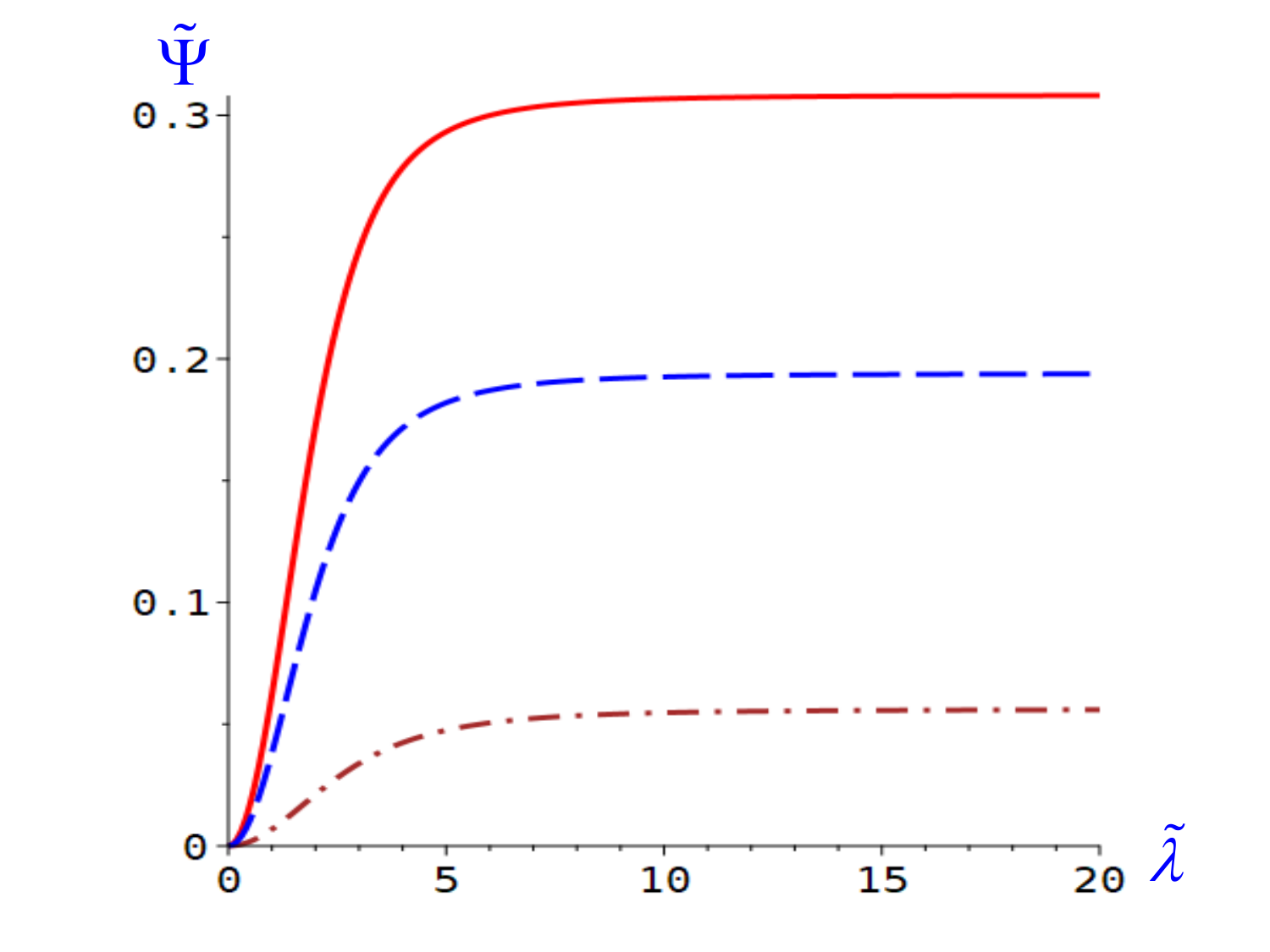}
    \caption{\n{F1} $\tilde{\Psi}$ as a function of $\tilde{\lambda}$ for the values $\tilde{r}_m=1.8$ (red solid line), $\tilde{r}_m=2.0$ (blue dashed line) and $\tilde{r}_m=3.0$ (brown dash and dotted line).}
\end{figure}

\begin{figure}[!hbt]
    \centering
      \includegraphics[width=0.4\textwidth]{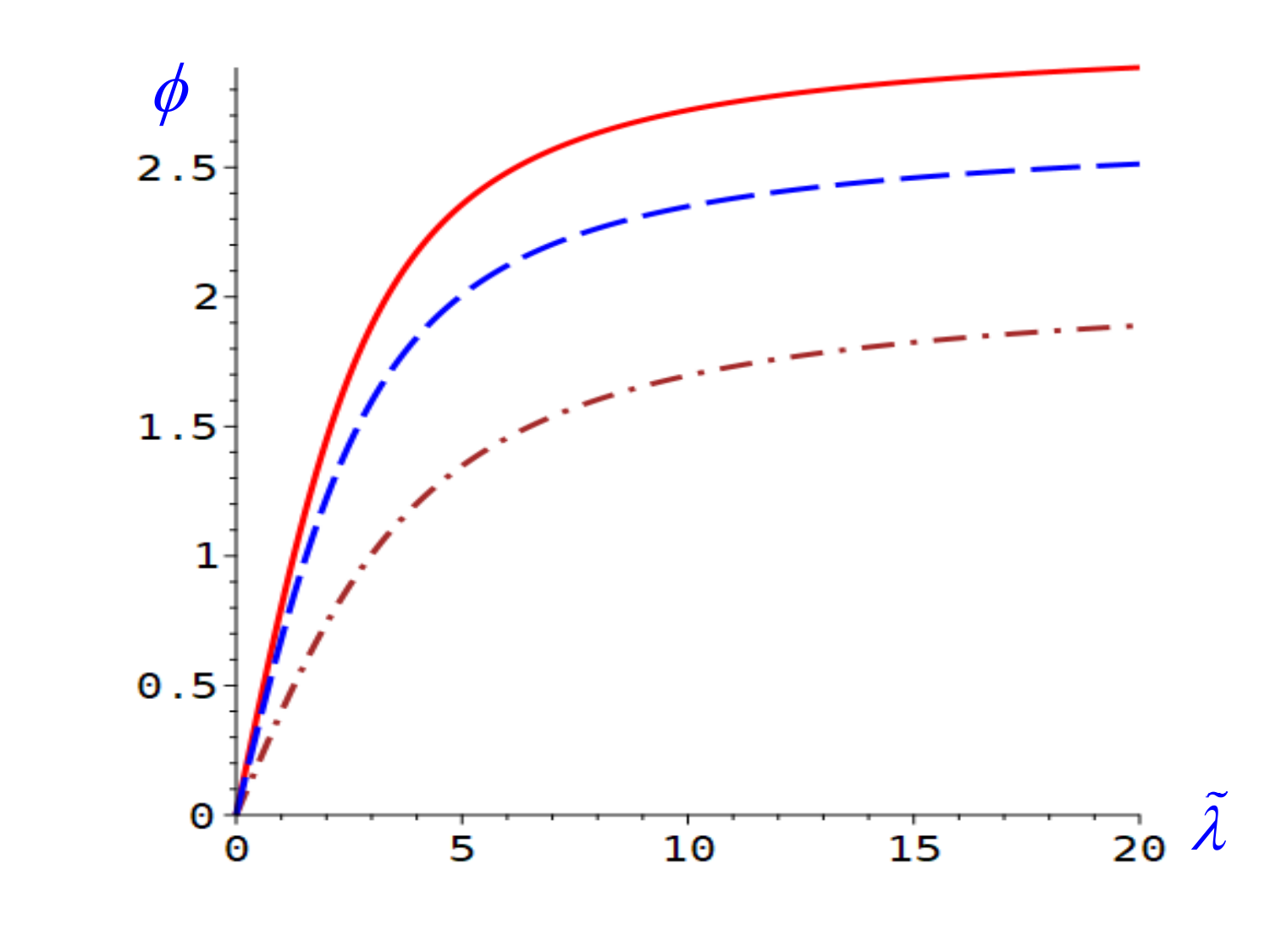}
    \caption{\n{F2} $\phi$ as a function of $\tilde{\lambda}$ for the values $\tilde{r}_m=1.8$ (red solid line), $\tilde{r}_m=2.0$ (blue dashed line) and $\tilde{r}_m=3.0$ (brown dash and dotted line).}
\end{figure}

\subsection{Tilting of orbit's plane}

Let us discuss now how a motion of the null ray changes when the one  "switches on" the interaction of the spin with the curvature. Let us assume  that at $\tilde{r}=\tilde{r}_m$ one has  $l^{\mu}=l_0^{\mu}$ and hence the perturbed ray lies in the equatorial plane  $\Pi_0$. As the result of the interaction of the helicity of the massless particle with the curvature its worldline for later and earlier time is shifted from this plane.  In the asymptotic domain $\Psi$ is practically constant and
the vectors $l_0^{\mu}$ and  $l^{\mu}$ has the following asymptotic form valid for $r\to \infty$
\be
\begin{split}
&l_0^{\mu}=(1,\vec{l}_0)\hh \vec{l}_0=\pm \vec{e}_{r}\, ,\\
&l^{\mu}=(1,\vec{l})\hh \vec{l}=\pm (\vec{e}_{r}+ \Psi_{\infty}\vec{e}_{\theta})\,
\end{split}
\ee
Here $\vec{e}_{r}$ and $\vec{e}_{\theta}$ are unit 3D vectors in the radial and $\theta$ directions, respectively.
A sign "minus" stands for the incoming ray, and sign "plus" stands for the outgoing one.

The incoming perturbed ray lies on the plane $\Pi_-$, while the outgoing ray lies on the plane $\Pi_+$. Both of these planes differ from $\Pi_0$. The shift parameter $\tilde{\Psi}_{\infty}$ determines the tilting angle between these asymptotic orbit's planes and the equatorial one.
This tilting angle $\Psi_{\infty}$ is
\be
\Psi_{\infty}=\dfrac{3}{2}\big(\dfrac{\lambdabar}{2M}\big)\tilde{\Psi}_{\infty}\, .
\ee
The quantity $\tilde{\Psi}_{\infty}$ is obtained by the integration of the first two equations in \eqref{eq3} with the initial conditions
\be
\tilde{\Psi}|_{\tilde{\lambda}=0}=0\hh \tilde{r}|_{\tilde{\lambda}=0}=\tilde{r}_m\, .
\ee

\begin{figure}[!hbt]
    \centering
      \includegraphics[width=0.4\textwidth]{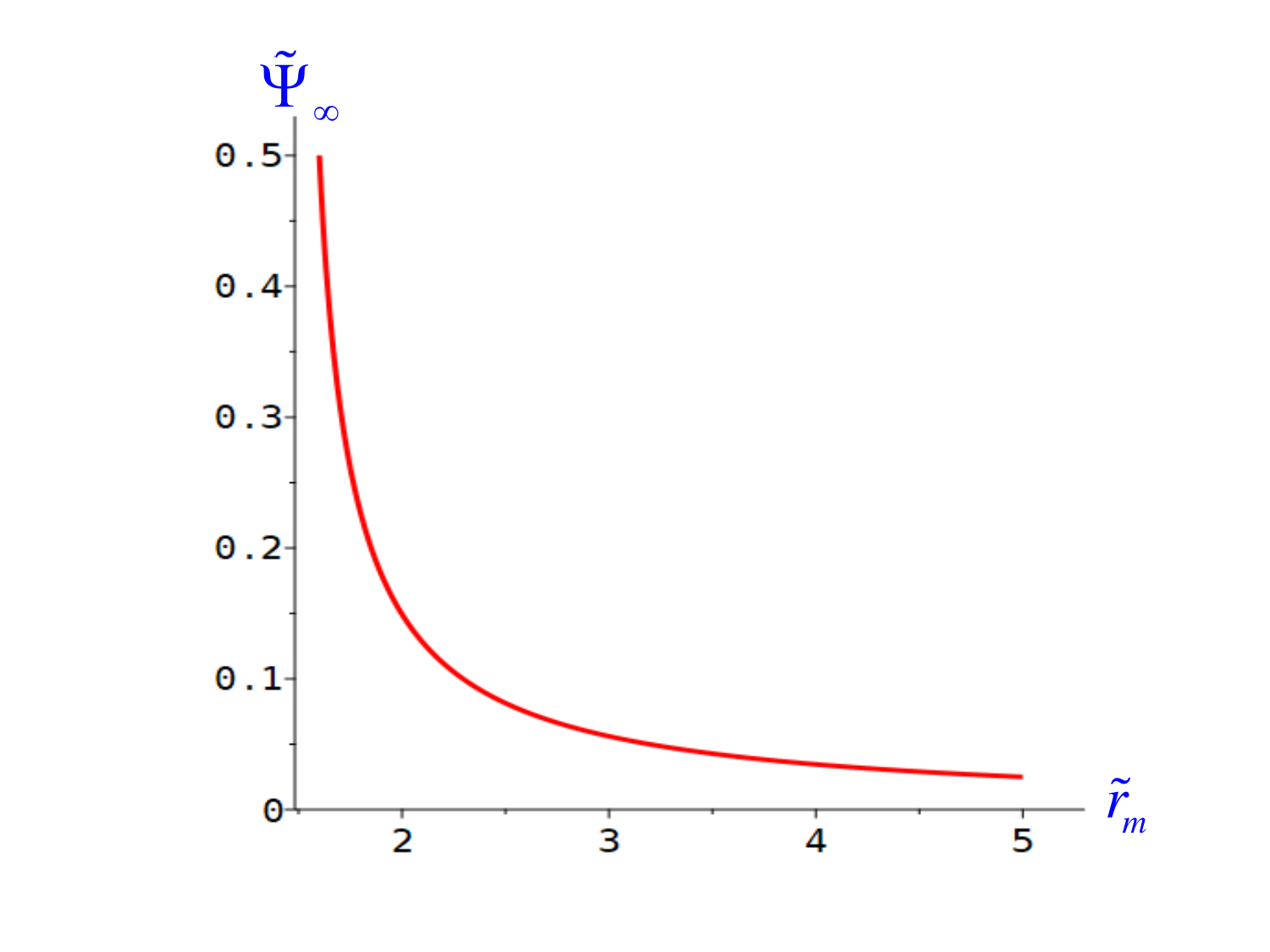}
    \caption{\n{F3} $\tilde{\Psi}_{\infty}$ as a function of $\tilde{r}_m$.}
\end{figure}

Performing this integration and taking the limit $\tilde{\lambda}\to \infty$ one gets $\tilde{\Psi}_{\infty}$ as a function of $\tilde{r}_m$. A plot of this function is shown at Fig.~\ref{F3}.  One can see that this function infinitely grows near $\tilde{r}_m=3/2$. This feature has a simple and natural explanation. In this regime a null ray spends some time revolving near the critical circular orbit and the tilting angle parameter is accumulated during this time.

\section{Discussion}

In this paper we discuss spinoptics of electromagnetic and weak gravitational waves in a spacetime of a static spherically symmetric black hole. We demonstrated that the spinoptics equations can be presented in a quite simple form which allows a detailed study. This became possible since the Schwarzschild geometry besides explicit symmetries generated by the Killing vectors also possesses a hidden symmetry. The generator of this hidden symmetry is a rank two closed conformal Killing-Yano tensor $h_{\mu\nu}$. It and its dual Killing-Yano tensor $k_{\mu\nu}$ allow one to construct two mutually orthogonal vectors which are parallel propagated  along a null geodesic. This result is used to obtain a complex null tetrad associated with the null geodesics, which are used in the spinoptics equations. Namely, a special projection of the Riemann curvature tensor on the complex null tetrad vectors determines the 4D "force" acting on the null ray. In the spinoptics approximation such a term describing the interaction of the photon's and graviton's spin with the curvature is proportional to $1/\omega$ and in the adopted high-frequency approximation is small. For this reason, one considers it as a perturbation. This implies that for the calculation of this term it is sufficient to use the complex null tetrads associated with the unperturbed null geodesic.

As a result, this "force" term can be obtained in an explicit analytical form. It is shown that in this approximation the corresponding "force" is directed orthogonal to the particle orbit's plane. The obtained equations describing this tilting orbit effect are integrated numerically. Let us emphasise that this  helicity dependent effect  is present in the field of a non-rotating black hole, and  it is quite different from the well known effect of the gravitational Faraday rotation induced by  the black hole rotation. The method of solving of the spinoptics equations described in this paper for the Schwarzschild geometry can be generalized to the case of the Kerr metric, describing  rotating  black holes.

\section*{Acknowledgements}
The author is grateful to the Yukawa Institute of Theoretical Physics of the Kyoto University for
its support and hospitality while this work was done. He also thanks Prof. Shinji Mukahyama and the members of the Gravity group for interesting and stimulating discussions during his stay as a visiting professor in the Yukawa Institute. This work was  supported by the Natural Sciences and
Engineering Research Council of Canada. The author is  also grateful to the Killam Trust for its financial support.


\begin{thebibliography}{44}%
\makeatletter
\providecommand \@ifxundefined [1]{%
 \@ifx{#1\undefined}
}%
\providecommand \@ifnum [1]{%
 \ifnum #1\expandafter \@firstoftwo
 \else \expandafter \@secondoftwo
 \fi
}%
\providecommand \@ifx [1]{%
 \ifx #1\expandafter \@firstoftwo
 \else \expandafter \@secondoftwo
 \fi
}%
\providecommand \natexlab [1]{#1}%
\providecommand \enquote  [1]{``#1''}%
\providecommand \bibnamefont  [1]{#1}%
\providecommand \bibfnamefont [1]{#1}%
\providecommand \citenamefont [1]{#1}%
\providecommand \href@noop [0]{\@secondoftwo}%
\providecommand \href [0]{\begingroup \@sanitize@url \@href}%
\providecommand \@href[1]{\@@startlink{#1}\@@href}%
\providecommand \@@href[1]{\endgroup#1\@@endlink}%
\providecommand \@@startlink[1]{}%
\providecommand \@@endlink[0]{}%
\providecommand \url  [0]{\begingroup\@sanitize@url \@url }%
\providecommand \@url [1]{\endgroup\@href {#1}{\urlprefix }}%
\providecommand \urlprefix  [0]{URL }%
\providecommand \Eprint [0]{\href }%
\providecommand \doibase [0]{http://dx.doi.org/}%
\providecommand \selectlanguage [0]{\@gobble}%
\providecommand \bibinfo  [0]{\@secondoftwo}%
\providecommand \bibfield  [0]{\@secondoftwo}%
\providecommand \translation [1]{[#1]}%
\providecommand \BibitemOpen [0]{}%
\providecommand \bibitemStop [0]{}%
\providecommand \bibitemNoStop [0]{.\EOS\space}%
\providecommand \EOS [0]{\spacefactor3000\relax}%
\providecommand \BibitemShut  [1]{\csname bibitem#1\endcsname}%
\let\auto@bib@innerbib\@empty
\bibitem [{\citenamefont {Debay}(1911)}]{Debay:1911}%
  \BibitemOpen
  \bibfield  {author} {\bibinfo {author} {\bibfnamefont {P.}~\bibnamefont
  {Debay}},\ }\href@noop {} {\bibfield  {journal} {\bibinfo  {journal} {Ann.
  Phys. (Leipzig)}\ }\textbf {\bibinfo {volume} {35}},\ \bibinfo {pages} {277}
  (\bibinfo {year} {1911})}\BibitemShut {NoStop}%
\bibitem [{\citenamefont {Misner}\ \emph {et~al.}(1974)\citenamefont {Misner},
  \citenamefont {Thorne},\ and\ \citenamefont {Wheeler}}]{MTW}%
  \BibitemOpen
  \bibfield  {author} {\bibinfo {author} {\bibfnamefont {C.~W.}\ \bibnamefont
  {Misner}}, \bibinfo {author} {\bibfnamefont {K.}~\bibnamefont {Thorne}}, \
  and\ \bibinfo {author} {\bibfnamefont {J.}~\bibnamefont {Wheeler}},\
  }\href@noop {} {\emph {\bibinfo {title} {{Gravitation}}}}\ (\bibinfo
  {publisher} {W.H. Freeman and Co., San Francisco},\ \bibinfo {year}
  {1974})\BibitemShut {NoStop}%
\bibitem [{\citenamefont {Skrotskii}(1957)}]{Skrotskii}%
  \BibitemOpen
  \bibfield  {author} {\bibinfo {author} {\bibfnamefont {G.~V.}\ \bibnamefont
  {Skrotskii}},\ }\bibfield  {title} {\enquote {\bibinfo {title} {{On the
  Influence of Gravity on the Light Propagation}},}\ }\href@noop {} {\bibfield
  {journal} {\bibinfo  {journal} {Soviet Phys. Doklady}\ }\textbf {\bibinfo
  {volume} {2}},\ \bibinfo {pages} {226} (\bibinfo {year} {1957})},\ \bibinfo
  {note} {[Akademia Nauk SSR, Doklady, {\bf 114}, 73, 1957]}\BibitemShut
  {NoStop}%
\bibitem [{\citenamefont {Plebanski}(1959)}]{Plebanski}%
  \BibitemOpen
  \bibfield  {author} {\bibinfo {author} {\bibfnamefont {J.}~\bibnamefont
  {Plebanski}},\ }\bibfield  {title} {\enquote {\bibinfo {title}
  {{Electromagnetic Waves in Gravitational Fields}},}\ }\href@noop {}
  {\bibfield  {journal} {\bibinfo  {journal} {Phys. Rev.}\ }\textbf {\bibinfo
  {volume} {118}},\ \bibinfo {pages} {1396} (\bibinfo {year}
  {1959})}\BibitemShut {NoStop}%
\bibitem [{\citenamefont {Godfrey}(1970)}]{God}%
  \BibitemOpen
  \bibfield  {author} {\bibinfo {author} {\bibfnamefont {B.~B.}\ \bibnamefont
  {Godfrey}},\ }\bibfield  {title} {\enquote {\bibinfo {title} {{Mach's
  Principle, the Kerr Metric, and Black-Hole Physics}},}\ }\href@noop {}
  {\bibfield  {journal} {\bibinfo  {journal} {Phys. Rev. D}\ }\textbf {\bibinfo
  {volume} {1}},\ \bibinfo {pages} {2721} (\bibinfo {year} {1970})}\BibitemShut
  {NoStop}%
\bibitem [{\citenamefont {Pineault}\ and\ \citenamefont {Roeder}(1977)}]{GF2}%
  \BibitemOpen
  \bibfield  {author} {\bibinfo {author} {\bibfnamefont {S.}~\bibnamefont
  {Pineault}}\ and\ \bibinfo {author} {\bibfnamefont {R.~C.}\ \bibnamefont
  {Roeder}},\ }\bibfield  {title} {\enquote {\bibinfo {title} {{Applications of
  Geometrical Optics to the Kerr Metric. I. Analytical Results}},}\ }\href@noop
  {} {\bibfield  {journal} {\bibinfo  {journal} {Astrophys. J.}\ }\textbf
  {\bibinfo {volume} {212}},\ \bibinfo {pages} {541} (\bibinfo {year}
  {1977})}\BibitemShut {NoStop}%
\bibitem [{\citenamefont {Connors}\ and\ \citenamefont {Stark}(1977)}]{GF3}%
  \BibitemOpen
  \bibfield  {author} {\bibinfo {author} {\bibfnamefont {P.~A.}\ \bibnamefont
  {Connors}}\ and\ \bibinfo {author} {\bibfnamefont {R.~F.}\ \bibnamefont
  {Stark}},\ }\bibfield  {title} {\enquote {\bibinfo {title} {{Observable
  gravitational effects on polarised radiation coming from near a black
  hole}},}\ }\href@noop {} {\bibfield  {journal} {\bibinfo  {journal} {Nature
  (London)}\ }\textbf {\bibinfo {volume} {269}},\ \bibinfo {pages} {128}
  (\bibinfo {year} {1977})}\BibitemShut {NoStop}%
\bibitem [{\citenamefont {Connors}\ \emph {et~al.}(1980)\citenamefont
  {Connors}, \citenamefont {Piran},\ and\ \citenamefont {Stark}}]{GF4}%
  \BibitemOpen
  \bibfield  {author} {\bibinfo {author} {\bibfnamefont {P.~A.}\ \bibnamefont
  {Connors}}, \bibinfo {author} {\bibfnamefont {T.}~\bibnamefont {Piran}}, \
  and\ \bibinfo {author} {\bibfnamefont {R.~F.}\ \bibnamefont {Stark}},\
  }\bibfield  {title} {\enquote {\bibinfo {title} {{Polarization features of
  X-ray radiation emitted near black holes}},}\ }\href@noop {} {\bibfield
  {journal} {\bibinfo  {journal} {Astrophys. J.}\ }\textbf {\bibinfo {volume}
  {235}},\ \bibinfo {pages} {224} (\bibinfo {year} {1980})}\BibitemShut
  {NoStop}%
\bibitem [{\citenamefont {Fayos}\ and\ \citenamefont {Llosa}(1982)}]{GF5}%
  \BibitemOpen
  \bibfield  {author} {\bibinfo {author} {\bibfnamefont {F.}~\bibnamefont
  {Fayos}}\ and\ \bibinfo {author} {\bibfnamefont {J.}~\bibnamefont {Llosa}},\
  }\bibfield  {title} {\enquote {\bibinfo {title} {{Gravitational Effects on
  the Polarization Plane}},}\ }\href@noop {} {\bibfield  {journal} {\bibinfo
  {journal} {General Relativity and Gravitation}\ }\textbf {\bibinfo {volume}
  {14}},\ \bibinfo {pages} {865} (\bibinfo {year} {1982})}\BibitemShut
  {NoStop}%
\bibitem [{\citenamefont {{Piran}}\ and\ \citenamefont
  {{Safier}}(1985)}]{Piran:1985}%
  \BibitemOpen
  \bibfield  {author} {\bibinfo {author} {\bibfnamefont {T.}~\bibnamefont
  {{Piran}}}\ and\ \bibinfo {author} {\bibfnamefont {P.~N.}\ \bibnamefont
  {{Safier}}},\ }\bibfield  {title} {\enquote {\bibinfo {title} {{A
  gravitational analogue of Faraday rotation}},}\ }\href@noop {} {\bibfield
  {journal} {\bibinfo  {journal} {\nat}\ }\textbf {\bibinfo {volume} {318}},\
  \bibinfo {pages} {271} (\bibinfo {year} {1985})}\BibitemShut {NoStop}%
\bibitem [{\citenamefont {Ishihara}\ \emph {et~al.}(1988)\citenamefont
  {Ishihara}, \citenamefont {Takahashi},\ and\ \citenamefont
  {Tomimatsu}}]{GF6}%
  \BibitemOpen
  \bibfield  {author} {\bibinfo {author} {\bibfnamefont {H.}~\bibnamefont
  {Ishihara}}, \bibinfo {author} {\bibfnamefont {M.}~\bibnamefont {Takahashi}},
  \ and\ \bibinfo {author} {\bibfnamefont {A.}~\bibnamefont {Tomimatsu}},\
  }\bibfield  {title} {\enquote {\bibinfo {title} {{Gravitational Faraday
  rotation induced by a Kerr black hole}},}\ }\href@noop {} {\bibfield
  {journal} {\bibinfo  {journal} {Phys.\ Rev.\ D}\ }\textbf {\bibinfo {volume}
  {38}},\ \bibinfo {pages} {472} (\bibinfo {year} {1988})}\BibitemShut
  {NoStop}%
\bibitem [{\citenamefont {Carini}\ \emph {et~al.}(1992)\citenamefont {Carini},
  \citenamefont {Feng}, \citenamefont {Li},\ and\ \citenamefont
  {Ruffini}}]{CariniRuffini}%
  \BibitemOpen
  \bibfield  {author} {\bibinfo {author} {\bibfnamefont {P.}~\bibnamefont
  {Carini}}, \bibinfo {author} {\bibfnamefont {L.~L.}\ \bibnamefont {Feng}},
  \bibinfo {author} {\bibfnamefont {M.}~\bibnamefont {Li}}, \ and\ \bibinfo
  {author} {\bibfnamefont {R.}~\bibnamefont {Ruffini}},\ }\bibfield  {title}
  {\enquote {\bibinfo {title} {{Phase evolution of the photon in Kerr
  spacetime}},}\ }\href@noop {} {\bibfield  {journal} {\bibinfo  {journal}
  {Phys.\ Rev.\ D}\ }\textbf {\bibinfo {volume} {46}},\ \bibinfo {pages} {5407}
  (\bibinfo {year} {1992})}\BibitemShut {NoStop}%
\bibitem [{\citenamefont {Perlick}\ and\ \citenamefont
  {Hasse}(1993)}]{Perlick_1993}%
  \BibitemOpen
  \bibfield  {author} {\bibinfo {author} {\bibfnamefont {V.}~\bibnamefont
  {Perlick}}\ and\ \bibinfo {author} {\bibfnamefont {W.}~\bibnamefont
  {Hasse}},\ }\bibfield  {title} {\enquote {\bibinfo {title} {{Gravitational
  Faraday effect in conformally stationary spacetimes}},}\ }\href {\doibase
  10.1088/0264-9381/10/1/015} {\bibfield  {journal} {\bibinfo  {journal}
  {Classical and Quantum Gravity}\ }\textbf {\bibinfo {volume} {10}},\ \bibinfo
  {pages} {147} (\bibinfo {year} {1993})}\BibitemShut {NoStop}%
\bibitem [{\citenamefont {Nouri-Zonoz}(1999)}]{GF7}%
  \BibitemOpen
  \bibfield  {author} {\bibinfo {author} {\bibfnamefont {M.}~\bibnamefont
  {Nouri-Zonoz}},\ }\bibfield  {title} {\enquote {\bibinfo {title}
  {{Gravitoelectromagnetic approach to the gravitational Faraday rotation in
  stationary space-times}},}\ }\href@noop {} {\bibfield  {journal} {\bibinfo
  {journal} {Phys.\ Rev.\ D}\ }\textbf {\bibinfo {volume} {60}},\ \bibinfo
  {pages} {024013} (\bibinfo {year} {1999})}\BibitemShut {NoStop}%
\bibitem [{\citenamefont {Sereno}(2004)}]{GF8}%
  \BibitemOpen
  \bibfield  {author} {\bibinfo {author} {\bibfnamefont {M.}~\bibnamefont
  {Sereno}},\ }\bibfield  {title} {\enquote {\bibinfo {title} {{Gravitational
  Faraday rotation in a weak gravitational field}},}\ }\href@noop {} {\bibfield
   {journal} {\bibinfo  {journal} {Phys.\ Rev.\ D}\ }\textbf {\bibinfo {volume}
  {69}},\ \bibinfo {pages} {087501} (\bibinfo {year} {2004})}\BibitemShut
  {NoStop}%
\bibitem [{\citenamefont {Sereno}(2005)}]{GF9}%
  \BibitemOpen
  \bibfield  {author} {\bibinfo {author} {\bibfnamefont {M.}~\bibnamefont
  {Sereno}},\ }\bibfield  {title} {\enquote {\bibinfo {title} {{Detecting
  gravitomagnetism with rotation of polarization by a gravitational lens}},}\
  }\href@noop {} {\bibfield  {journal} {\bibinfo  {journal} {Mon. Not. R.
  Astron. Soc.}\ }\textbf {\bibinfo {volume} {356}},\ \bibinfo {pages} {381}
  (\bibinfo {year} {2005})}\BibitemShut {NoStop}%
\bibitem [{\citenamefont {Halilsoy}\ and\ \citenamefont
  {Gurtug}(2007)}]{Halilsoy:2006ev}%
  \BibitemOpen
  \bibfield  {author} {\bibinfo {author} {\bibfnamefont {M.}~\bibnamefont
  {Halilsoy}}\ and\ \bibinfo {author} {\bibfnamefont {O.}~\bibnamefont
  {Gurtug}},\ }\bibfield  {title} {\enquote {\bibinfo {title} {{Search for
  gravitational waves through the electromagnetic Faraday rotation}},}\ }\href
  {\doibase 10.1103/PhysRevD.75.124021} {\bibfield  {journal} {\bibinfo
  {journal} {Phys. Rev. D}\ }\textbf {\bibinfo {volume} {75}},\ \bibinfo
  {pages} {124021} (\bibinfo {year} {2007})},\ \Eprint
  {http://arxiv.org/abs/gr-qc/0612107}{arXiv:gr-qc/0612107}\BibitemShut
  {NoStop}%
\bibitem [{\citenamefont {Brodutch}\ \emph {et~al.}(2011)\citenamefont
  {Brodutch}, \citenamefont {Demarie},\ and\ \citenamefont {Terno}}]{GF10}%
  \BibitemOpen
  \bibfield  {author} {\bibinfo {author} {\bibfnamefont {A.}~\bibnamefont
  {Brodutch}}, \bibinfo {author} {\bibfnamefont {T.~F.}\ \bibnamefont
  {Demarie}}, \ and\ \bibinfo {author} {\bibfnamefont {D.~R.}\ \bibnamefont
  {Terno}},\ }\bibfield  {title} {\enquote {\bibinfo {title} {{Photon
  polarization and geometric phase in general relativity}},}\ }\href@noop {}
  {\bibfield  {journal} {\bibinfo  {journal} {Phys.\ Rev.\ D}\ }\textbf
  {\bibinfo {volume} {84}},\ \bibinfo {pages} {104043} (\bibinfo {year}
  {2011})}\BibitemShut {NoStop}%
\bibitem [{\citenamefont {Frolov}\ and\ \citenamefont
  {Shoom}(2011)}]{Frolov:2011mh}%
  \BibitemOpen
  \bibfield  {author} {\bibinfo {author} {\bibfnamefont {V.~P.}\ \bibnamefont
  {Frolov}}\ and\ \bibinfo {author} {\bibfnamefont {A.~A.}\ \bibnamefont
  {Shoom}},\ }\bibfield  {title} {\enquote {\bibinfo {title} {{Spinoptics in a
  stationary spacetime}},}\ }\href {\doibase 10.1103/PhysRevD.84.044026}
  {\bibfield  {journal} {\bibinfo  {journal} {Phys. Rev. D}\ }\textbf {\bibinfo
  {volume} {84}},\ \bibinfo {pages} {044026} (\bibinfo {year} {2011})},\
  \Eprint {http://arxiv.org/abs/1105.5629}{arXiv:1105.5629 [gr-qc]}\BibitemShut
  {NoStop}%
\bibitem [{\citenamefont {Ghosh}\ and\ \citenamefont {Sen}(2016)}]{Ghosh}%
  \BibitemOpen
  \bibfield  {author} {\bibinfo {author} {\bibfnamefont {T.}~\bibnamefont
  {Ghosh}}\ and\ \bibinfo {author} {\bibfnamefont {A.~K.}\ \bibnamefont
  {Sen}},\ }\bibfield  {title} {\enquote {\bibinfo {title} {{The Effect of
  Gravitation on the Polarization State of a Light ray}},}\ }\href@noop {}
  {\bibfield  {journal} {\bibinfo  {journal} {Astrophys.\ J.}\ }\textbf
  {\bibinfo {volume} {833}},\ \bibinfo {pages} {82} (\bibinfo {year}
  {2016})}\BibitemShut {NoStop}%
\bibitem [{\citenamefont {Dolan}(2017)}]{Dolan:2018nzc}%
  \BibitemOpen
  \bibfield  {author} {\bibinfo {author} {\bibfnamefont {S.~R.}\ \bibnamefont
  {Dolan}},\ }\bibfield  {title} {\enquote {\bibinfo {title} {{Geometrical
  optics for scalar, electromagnetic and gravitational waves on curved
  spacetime}},}\ }\href {\doibase 10.1142/S0218271818430101} {\bibfield
  {journal} {\bibinfo  {journal} {Int. J. Mod. Phys. D}\ }\textbf {\bibinfo
  {volume} {27}},\ \bibinfo {pages} {1843010} (\bibinfo {year} {2017})},\
  \Eprint {http://arxiv.org/abs/1806.08617}{arXiv:1806.08617
  [gr-qc]}\BibitemShut {NoStop}%
\bibitem [{\citenamefont {Dolan}(2018)}]{Dolan:2018ydp}%
  \BibitemOpen
  \bibfield  {author} {\bibinfo {author} {\bibfnamefont {S.~R.}\ \bibnamefont
  {Dolan}},\ }\bibfield  {title} {\enquote {\bibinfo {title} {{Higher-order
  geometrical optics for electromagnetic waves on a curved spacetime}},}\
  }\href@noop {} {\  (\bibinfo {year} {2018})},\ \Eprint
  {http://arxiv.org/abs/1801.02273}{arXiv:1801.02273 [gr-qc]}\BibitemShut
  {NoStop}%
\bibitem [{\citenamefont {Hou}\ \emph {et~al.}(2019)\citenamefont {Hou},
  \citenamefont {Fan},\ and\ \citenamefont {Zhu}}]{Hou:2019wdg}%
  \BibitemOpen
  \bibfield  {author} {\bibinfo {author} {\bibfnamefont {S.}~\bibnamefont
  {Hou}}, \bibinfo {author} {\bibfnamefont {X.-L.}\ \bibnamefont {Fan}}, \ and\
  \bibinfo {author} {\bibfnamefont {Z.-H.}\ \bibnamefont {Zhu}},\ }\bibfield
  {title} {\enquote {\bibinfo {title} {{Gravitational Lensing of Gravitational
  Waves: Rotation of Polarization Plane}},}\ }\href {\doibase
  10.1103/PhysRevD.100.064028} {\bibfield  {journal} {\bibinfo  {journal}
  {Phys. Rev. D}\ }\textbf {\bibinfo {volume} {100}},\ \bibinfo {pages}
  {064028} (\bibinfo {year} {2019})},\ \Eprint
  {http://arxiv.org/abs/1907.07486}{arXiv:1907.07486 [gr-qc]}\BibitemShut
  {NoStop}%
\bibitem [{\citenamefont {Li}\ \emph {et~al.}(2022)\citenamefont {Li},
  \citenamefont {Qiao}, \citenamefont {Zhao},\ and\ \citenamefont
  {Er}}]{Li:2022izh}%
  \BibitemOpen
  \bibfield  {author} {\bibinfo {author} {\bibfnamefont {Z.}~\bibnamefont
  {Li}}, \bibinfo {author} {\bibfnamefont {J.}~\bibnamefont {Qiao}}, \bibinfo
  {author} {\bibfnamefont {W.}~\bibnamefont {Zhao}}, \ and\ \bibinfo {author}
  {\bibfnamefont {X.}~\bibnamefont {Er}},\ }\bibfield  {title} {\enquote
  {\bibinfo {title} {{Gravitational Faraday Rotation of gravitational waves by
  a Kerr black hole}},}\ }\href {\doibase 10.1088/1475-7516/2022/10/095}
  {\bibfield  {journal} {\bibinfo  {journal} {JCAP}\ }\textbf {\bibinfo
  {volume} {10}},\ \bibinfo {pages} {095} (\bibinfo {year} {2022})},\ \Eprint
  {http://arxiv.org/abs/2204.10512}{arXiv:2204.10512 [gr-qc]}\BibitemShut
  {NoStop}%
\bibitem [{\citenamefont {Shoom}(2022)}]{Shoom:2022oer}%
  \BibitemOpen
  \bibfield  {author} {\bibinfo {author} {\bibfnamefont {A.~A.}\ \bibnamefont
  {Shoom}},\ }\bibfield  {title} {\enquote {\bibinfo {title} {{Faraday effect
  of light caused by plane gravitational wave}},}\ }\href@noop {} {\  (\bibinfo
  {year} {2022})},\ \Eprint {http://arxiv.org/abs/2206.08867}{arXiv:2206.08867
  [gr-qc]}\BibitemShut {NoStop}%
\bibitem [{\citenamefont {{Mathisson}}(2010)}]{MATHISSON}%
  \BibitemOpen
  \bibfield  {author} {\bibinfo {author} {\bibfnamefont {M.}~\bibnamefont
  {{Mathisson}}},\ }\bibfield  {title} {\enquote {\bibinfo {title}
  {{Republication of: New mechanics of material systems}},}\ }\href {\doibase
  10.1007/s10714-010-0939-y} {\bibfield  {journal} {\bibinfo  {journal}
  {General Relativity and Gravitation}\ }\textbf {\bibinfo {volume} {42}},\
  \bibinfo {pages} {1011} (\bibinfo {year} {2010})}\BibitemShut {NoStop}%
\bibitem [{\citenamefont {{Papapetrou}}(1951)}]{PAPAPETROU}%
  \BibitemOpen
  \bibfield  {author} {\bibinfo {author} {\bibfnamefont {A.}~\bibnamefont
  {{Papapetrou}}},\ }\bibfield  {title} {\enquote {\bibinfo {title} {{Spinning
  Test-Particles in General Relativity. I}},}\ }\href {\doibase
  10.1098/rspa.1951.0200} {\bibfield  {journal} {\bibinfo  {journal}
  {Proceedings of the Royal Society of London Series A}\ }\textbf {\bibinfo
  {volume} {209}},\ \bibinfo {pages} {248} (\bibinfo {year}
  {1951})}\BibitemShut {NoStop}%
\bibitem [{\citenamefont {{Dixon}}(1970)}]{DIXON}%
  \BibitemOpen
  \bibfield  {author} {\bibinfo {author} {\bibfnamefont {W.~G.}\ \bibnamefont
  {{Dixon}}},\ }\bibfield  {title} {\enquote {\bibinfo {title} {{Dynamics of
  Extended Bodies in General Relativity. I. Momentum and Angular Momentum}},}\
  }\href {\doibase 10.1098/rspa.1970.0020} {\bibfield  {journal} {\bibinfo
  {journal} {Proceedings of the Royal Society of London Series A}\ }\textbf
  {\bibinfo {volume} {314}},\ \bibinfo {pages} {499} (\bibinfo {year}
  {1970})}\BibitemShut {NoStop}%
\bibitem [{\citenamefont {Frolov}\ and\ \citenamefont
  {Shoom}(2012)}]{Frolov:2012zn}%
  \BibitemOpen
  \bibfield  {author} {\bibinfo {author} {\bibfnamefont {V.~P.}\ \bibnamefont
  {Frolov}}\ and\ \bibinfo {author} {\bibfnamefont {A.~A.}\ \bibnamefont
  {Shoom}},\ }\bibfield  {title} {\enquote {\bibinfo {title} {{Scattering of
  circularly polarized light by a rotating black hole}},}\ }\href {\doibase
  10.1103/PhysRevD.86.024010} {\bibfield  {journal} {\bibinfo  {journal} {Phys.
  Rev. D}\ }\textbf {\bibinfo {volume} {86}},\ \bibinfo {pages} {024010}
  (\bibinfo {year} {2012})},\ \Eprint
  {http://arxiv.org/abs/1205.4479}{arXiv:1205.4479 [gr-qc]}\BibitemShut
  {NoStop}%
\bibitem [{\citenamefont {Yoo}(2012)}]{Yoo:2012vv}%
  \BibitemOpen
  \bibfield  {author} {\bibinfo {author} {\bibfnamefont {C.-M.}\ \bibnamefont
  {Yoo}},\ }\bibfield  {title} {\enquote {\bibinfo {title} {{Notes on
  Spinoptics in a Stationary Spacetime}},}\ }\href {\doibase
  10.1103/PhysRevD.86.084005} {\bibfield  {journal} {\bibinfo  {journal} {Phys.
  Rev. D}\ }\textbf {\bibinfo {volume} {86}},\ \bibinfo {pages} {084005}
  (\bibinfo {year} {2012})},\ \Eprint
  {http://arxiv.org/abs/1207.6833}{arXiv:1207.6833 [gr-qc]}\BibitemShut
  {NoStop}%
\bibitem [{\citenamefont {Oancea}\ \emph {et~al.}(2019)\citenamefont {Oancea},
  \citenamefont {Paganini}, \citenamefont {Joudioux},\ and\ \citenamefont
  {Andersson}}]{Oancea:2019pgm}%
  \BibitemOpen
  \bibfield  {author} {\bibinfo {author} {\bibfnamefont {M.~A.}\ \bibnamefont
  {Oancea}}, \bibinfo {author} {\bibfnamefont {C.~F.}\ \bibnamefont
  {Paganini}}, \bibinfo {author} {\bibfnamefont {J.}~\bibnamefont {Joudioux}},
  \ and\ \bibinfo {author} {\bibfnamefont {L.}~\bibnamefont {Andersson}},\
  }\bibfield  {title} {\enquote {\bibinfo {title} {{An overview of the
  gravitational spin Hall effect}},}\ }\href@noop {} {\  (\bibinfo {year}
  {2019})},\ \Eprint {http://arxiv.org/abs/1904.09963}{arXiv:1904.09963
  [gr-qc]}\BibitemShut {NoStop}%
\bibitem [{\citenamefont {Oancea}\ \emph {et~al.}(2020)\citenamefont {Oancea},
  \citenamefont {Joudioux}, \citenamefont {Dodin}, \citenamefont {Ruiz},
  \citenamefont {Paganini},\ and\ \citenamefont {Andersson}}]{Oancea:2020khc}%
  \BibitemOpen
  \bibfield  {author} {\bibinfo {author} {\bibfnamefont {M.~A.}\ \bibnamefont
  {Oancea}}, \bibinfo {author} {\bibfnamefont {J.}~\bibnamefont {Joudioux}},
  \bibinfo {author} {\bibfnamefont {I.}~\bibnamefont {Dodin}}, \bibinfo
  {author} {\bibfnamefont {D.}~\bibnamefont {Ruiz}}, \bibinfo {author}
  {\bibfnamefont {C.~F.}\ \bibnamefont {Paganini}}, \ and\ \bibinfo {author}
  {\bibfnamefont {L.}~\bibnamefont {Andersson}},\ }\bibfield  {title} {\enquote
  {\bibinfo {title} {{The gravitational spin Hall effect of light}},}\
  }\href@noop {} {\  (\bibinfo {year} {2020})},\ \Eprint
  {http://arxiv.org/abs/2003.04553}{arXiv:2003.04553 [gr-qc]}\BibitemShut
  {NoStop}%
\bibitem [{\citenamefont {Frolov}(2020)}]{Frolov:2020uhn}%
  \BibitemOpen
  \bibfield  {author} {\bibinfo {author} {\bibfnamefont {V.~P.}\ \bibnamefont
  {Frolov}},\ }\bibfield  {title} {\enquote {\bibinfo {title} {{Maxwell
  equations in a curved spacetime: Spin optics approximation}},}\ }\href
  {\doibase 10.1103/PhysRevD.102.084013} {\bibfield  {journal} {\bibinfo
  {journal} {Phys. Rev. D}\ }\textbf {\bibinfo {volume} {102}},\ \bibinfo
  {pages} {084013} (\bibinfo {year} {2020})},\ \Eprint
  {http://arxiv.org/abs/2007.03743}{arXiv:2007.03743 [gr-qc]}\BibitemShut
  {NoStop}%
\bibitem [{\citenamefont {Dahal}(2023)}]{Dahal:2022gop}%
  \BibitemOpen
  \bibfield  {author} {\bibinfo {author} {\bibfnamefont {P.~K.}\ \bibnamefont
  {Dahal}},\ }\bibfield  {title} {\enquote {\bibinfo {title} {{Covariant
  formulation of spin optics for electromagnetic waves}},}\ }\href {\doibase
  10.1007/s00340-022-07952-2} {\bibfield  {journal} {\bibinfo  {journal} {Appl.
  Phys. B}\ }\textbf {\bibinfo {volume} {129}},\ \bibinfo {pages} {11}
  (\bibinfo {year} {2023})},\ \Eprint
  {http://arxiv.org/abs/2208.04725}{arXiv:2208.04725 [gr-qc]}\BibitemShut
  {NoStop}%
\bibitem [{\citenamefont {Yamamoto}(2018)}]{Yamamoto:2017gla}%
  \BibitemOpen
  \bibfield  {author} {\bibinfo {author} {\bibfnamefont {N.}~\bibnamefont
  {Yamamoto}},\ }\bibfield  {title} {\enquote {\bibinfo {title} {{Spin Hall
  effect of gravitational waves}},}\ }\href {\doibase
  10.1103/PhysRevD.98.061701} {\bibfield  {journal} {\bibinfo  {journal} {Phys.
  Rev. D}\ }\textbf {\bibinfo {volume} {98}},\ \bibinfo {pages} {061701}
  (\bibinfo {year} {2018})},\ \Eprint
  {http://arxiv.org/abs/1708.03113}{arXiv:1708.03113 [hep-th]}\BibitemShut
  {NoStop}%
\bibitem [{\citenamefont {Andersson}\ \emph {et~al.}(2021)\citenamefont
  {Andersson}, \citenamefont {Joudioux}, \citenamefont {Oancea},\ and\
  \citenamefont {Raj}}]{Andersson:2020gsj}%
  \BibitemOpen
  \bibfield  {author} {\bibinfo {author} {\bibfnamefont {L.}~\bibnamefont
  {Andersson}}, \bibinfo {author} {\bibfnamefont {J.}~\bibnamefont {Joudioux}},
  \bibinfo {author} {\bibfnamefont {M.~A.}\ \bibnamefont {Oancea}}, \ and\
  \bibinfo {author} {\bibfnamefont {A.}~\bibnamefont {Raj}},\ }\bibfield
  {title} {\enquote {\bibinfo {title} {{Propagation of polarized gravitational
  waves}},}\ }\href {\doibase 10.1103/PhysRevD.103.044053} {\bibfield
  {journal} {\bibinfo  {journal} {Phys. Rev. D}\ }\textbf {\bibinfo {volume}
  {103}},\ \bibinfo {pages} {044053} (\bibinfo {year} {2021})},\ \Eprint
  {http://arxiv.org/abs/2012.08363}{arXiv:2012.08363 [gr-qc]}\BibitemShut
  {NoStop}%
\bibitem [{\citenamefont {Dahal}(2022)}]{Dahal:2021qel}%
  \BibitemOpen
  \bibfield  {author} {\bibinfo {author} {\bibfnamefont {P.~K.}\ \bibnamefont
  {Dahal}},\ }\bibfield  {title} {\enquote {\bibinfo {title} {{Spin Optics for
  Gravitational Waves}},}\ }\href {\doibase 10.3390/astronomy1030016}
  {\bibfield  {journal} {\bibinfo  {journal} {Astronomy}\ }\textbf {\bibinfo
  {volume} {1}},\ \bibinfo {pages} {271} (\bibinfo {year} {2022})},\ \Eprint
  {http://arxiv.org/abs/2107.02761}{arXiv:2107.02761 [gr-qc]}\BibitemShut
  {NoStop}%
\bibitem [{\citenamefont {Kubota}\ \emph {et~al.}(2024)\citenamefont {Kubota},
  \citenamefont {Arai},\ and\ \citenamefont {Mukohyama}}]{Kubota:2023dlz}%
  \BibitemOpen
  \bibfield  {author} {\bibinfo {author} {\bibfnamefont {K.-i.}\ \bibnamefont
  {Kubota}}, \bibinfo {author} {\bibfnamefont {S.}~\bibnamefont {Arai}}, \ and\
  \bibinfo {author} {\bibfnamefont {S.}~\bibnamefont {Mukohyama}},\ }\bibfield
  {title} {\enquote {\bibinfo {title} {{Spin optics for gravitational waves
  lensed by a rotating object}},}\ }\href {\doibase
  10.1103/PhysRevD.109.044027} {\bibfield  {journal} {\bibinfo  {journal}
  {Phys. Rev. D}\ }\textbf {\bibinfo {volume} {109}},\ \bibinfo {pages}
  {044027} (\bibinfo {year} {2024})},\ \Eprint
  {http://arxiv.org/abs/2309.11024}{arXiv:2309.11024 [gr-qc]}\BibitemShut
  {NoStop}%
\bibitem [{\citenamefont {Frolov}(2024)}]{Frolov:2024ebe}%
  \BibitemOpen
  \bibfield  {author} {\bibinfo {author} {\bibfnamefont {V.~P.}\ \bibnamefont
  {Frolov}},\ }\bibfield  {title} {\enquote {\bibinfo {title} {{Spinoptics in a
  curved spacetime}},}\ }\href {\doibase 10.1103/PhysRevD.110.064020}
  {\bibfield  {journal} {\bibinfo  {journal} {Phys. Rev. D}\ }\textbf {\bibinfo
  {volume} {110}},\ \bibinfo {pages} {064020} (\bibinfo {year} {2024})},\
  \Eprint {http://arxiv.org/abs/2405.01777}{arXiv:2405.01777
  [gr-qc]}\BibitemShut {NoStop}%
\bibitem [{\citenamefont {Frolov}\ and\ \citenamefont
  {Shoom}(2024)}]{Frolov:2024qow}%
  \BibitemOpen
  \bibfield  {author} {\bibinfo {author} {\bibfnamefont {V.~P.}\ \bibnamefont
  {Frolov}}\ and\ \bibinfo {author} {\bibfnamefont {A.~A.}\ \bibnamefont
  {Shoom}},\ }\bibfield  {title} {\enquote {\bibinfo {title} {{Gravitational
  spinoptics in a curved space-time}},}\ }\href@noop {} {\  (\bibinfo {year}
  {2024})},\ \Eprint {http://arxiv.org/abs/2406.17905}{arXiv:2406.17905
  [gr-qc]}\BibitemShut {NoStop}%
\bibitem [{\citenamefont {Frolov}\ and\ \citenamefont
  {Zelnikov}(2011)}]{FRO_ZEL}%
  \BibitemOpen
  \bibfield  {author} {\bibinfo {author} {\bibfnamefont {V.~P.}\ \bibnamefont
  {Frolov}}\ and\ \bibinfo {author} {\bibfnamefont {A.}~\bibnamefont
  {Zelnikov}},\ }\href@noop {} {\emph {\bibinfo {title} {{Introduction to Black
  Hole Physics}}}}\ (\bibinfo  {publisher} {Oxford University Press},\ \bibinfo
  {year} {2011})\BibitemShut {NoStop}%
\bibitem [{\citenamefont {Marck}(1983{\natexlab{a}})}]{MARCK_2}%
  \BibitemOpen
  \bibfield  {author} {\bibinfo {author} {\bibfnamefont {J.-A.}\ \bibnamefont
  {Marck}},\ }\bibfield  {title} {\enquote {\bibinfo {title} {Solution to the
  equations of parallel transport in kerr geometry; tidal tensor},}\ }\href
  {http://www.jstor.org/stable/2397341} {\bibfield  {journal} {\bibinfo
  {journal} {Proceedings of the Royal Society of London. Series A, Mathematical
  and Physical Sciences}\ }\textbf {\bibinfo {volume} {385}},\ \bibinfo {pages}
  {431} (\bibinfo {year} {1983}{\natexlab{a}})}\BibitemShut {NoStop}%
\bibitem [{\citenamefont {Marck}(1983{\natexlab{b}})}]{MARCK_1}%
  \BibitemOpen
  \bibfield  {author} {\bibinfo {author} {\bibfnamefont {J.-A.}\ \bibnamefont
  {Marck}},\ }\bibfield  {title} {\enquote {\bibinfo {title} {Parallel-tetrad
  on null geodesics in kerr-newman space-time},}\ }\href {\doibase
  https://doi.org/10.1016/0375-9601(83)90197-4} {\bibfield  {journal} {\bibinfo
   {journal} {Physics Letters A}\ }\textbf {\bibinfo {volume} {97}},\ \bibinfo
  {pages} {140} (\bibinfo {year} {1983}{\natexlab{b}})}\BibitemShut {NoStop}%
\bibitem [{\citenamefont {Frolov}\ \emph {et~al.}(2017)\citenamefont {Frolov},
  \citenamefont {Krtous},\ and\ \citenamefont {Kubiznak}}]{LIVING}%
  \BibitemOpen
  \bibfield  {author} {\bibinfo {author} {\bibfnamefont {V.~P.}\ \bibnamefont
  {Frolov}}, \bibinfo {author} {\bibfnamefont {P.}~\bibnamefont {Krtous}}, \
  and\ \bibinfo {author} {\bibfnamefont {D.}~\bibnamefont {Kubiznak}},\
  }\bibfield  {title} {\enquote {\bibinfo {title} {{Black holes, hidden
  symmetries, and complete integrability}},}\ }\href {\doibase
  10.1007/s41114-017-0009-9} {\bibfield  {journal} {\bibinfo  {journal} {Living
  Rev. Rel.}\ }\textbf {\bibinfo {volume} {20}},\ \bibinfo {pages} {6}
  (\bibinfo {year} {2017})},\ \Eprint
  {http://arxiv.org/abs/1705.05482}{arXiv:1705.05482 [gr-qc]}\BibitemShut
  {NoStop}%
\end{thebibliography}

%

\end{document}